\documentclass[11pt,twoside]{article}
\usepackage{asp2010}

\resetcounters

\begin{document}

\title{Vortices in the solar photosphere}
\author{S.~Shelyag$^1$, V.~Fedun$^2$, R.~Erd{\'e}lyi$^2$, F.P.~Keenan$^1$, and M.~Mathioudakis$^1$
\affil{$^1$Astrophysics Research Centre, Department of Mathematics and Physics, Queen's University, Belfast, BT7 1NN, UK}
\affil{$^2$Solar Physics and Space Plasma Research Centre (SP$^2$RC), Department of Applied Mathematics, 
University of Sheffield, Sheffield, S3 7RH, UK}
}

\begin{abstract}
Using numerical simulations of the magnetised solar photosphere and radiative diagnostics of the simulated photospheric models,
we further analyse the physical nature of magnetic photospheric intergranular vortices.  We confirm the magnetic nature of the 
vortices and find that most MHD Umov-Poynting flux is produced by horizontal vortex motions in the magnetised intergranular 
lanes. In addition, we consider possible ways to directly observe photospheric magnetic vortices using spectropolarimetry.  
Although horizontal plasma motions cannot be detected in the spectropolarimetric observations of solar disk centre, we find an 
observational signature of photospheric vortices in simulated observations of Stokes-$V$ amplitude asymmetry close to the 
solar limb. Potential ways to find the vortices in the observations are discussed.
\end{abstract}

\section{Introduction}
The recent discovery of vortex motions in the solar photosphere has raised the question of the role of photospheric 
vortices in the energy transport from the solar interior to the outer atmosphere. \citet{bonet1} showed that Magnetic Bright Points (MBPs)  
are subject to vortex motions in the intergranular lanes. An analysis of small-scale intergranular vortex 
structures and their motions has been presented by \citet{vargas1}. High-resolution multi-wavelength observations 
\citep{wedemeyer1} suggest that the vortices, corresponding to MBP motions, coincide spatially with 
chromospheric swirls. \citet{moll1} reported on the formation of small-scale intergranular vortices in  weak 
magnetic field simulations of the solar photosphere. \citet{shelyag2011a,shelyag2011b} have demonstrated 
that solar atmospheric vorticity generation, which reveals itself as both MBP rotary motions and upper-atmospheric 
swirls, efficiently occurs only in intergranular magnetic field concentrations, and the vorticity generation in the upper 
solar photosphere is essentially a magnetic phenomenon. Vortex motions in small-scale 
magnetic field concentrations can generate a variety of magneto-hydrodynamic waves \citep{erdelyi1,jess1,fedun2011}, 
which propagate to the outer layers of solar atmosphere, where the energy can be 
dissipated \citep[see e.g.][]{zaqarashvili1}. This mechanism may be responsible for the transport of energy 
from the solar interior towards the higher layers of the solar atmosphere, including the corona. 

This paper is devoted to some further analysis of photospheric magnetic vortices. We demonstrate 
the generation of positive, directed outwards Umov-Poynting flux and also show that the vertical component of the 
Umov-Poynting flux vector is generated mainly
by horizontal vortex motions. This finding is confirmed by the numerical simulation of an individual cylindrically
symmetric magnetic flux tube embedded in a quiet convectively stable atmospheric model. We also perform
detailed spectropolarimetric diagnostics with the FeI $630.2~\mathrm{nm}$ line to explore 
possibilities for observing intergranular vortex motions close to the solar limb.

The structure of this paper is organised as follows. In Section 2, we briefly describe our simulation 
setup and show the mechanisms to produce magnetic photospheric vorticity. We demonstrate the presence 
of Umov-Poynting flux connected to photospheric vortices in Section 3, and further support this evidence by
performing a test flux tube simulation in Section 4, and discuss possibilities for observing photospheric vorticity 
in Section 5. Our conclusions are given in Section 6.

\section{Vorticity generation is the solar photosphere}
We use the MURaM code \citep{voegler1} to undertake simulations of the magnetised photosphere. The code
is extensively used in the solar physics community and is described in great detail in the literature 
\citep[see e.g.][]{shelyagbp2,rempel1}. Our simulation procedure is similar to that 
described in \citet{shelyag2011a}, and here we provide only a brief description. The numerical domain 
is $12 \times 12 \times 1.4~\mathrm{Mm^3}$ size, resolved by $480 \times 480 \times 100$ grid cells. 
A uniform magnetic field of $200~\mathrm{G}$ strength is introduced into a well-developed, non-magnetic 
photospheric convection snapshot, and a sequence of snapshots, containing the photospheric plasma
parameters, is recorded. Here we use a well-developed snapshot of 
photospheric magneto-convection recorded after 1 hour of the physical simulation time, during which
the magnetic field has collapsed into the intergranular lanes and the simulation has achieved statistical
stability.

\begin{figure}
\plotone{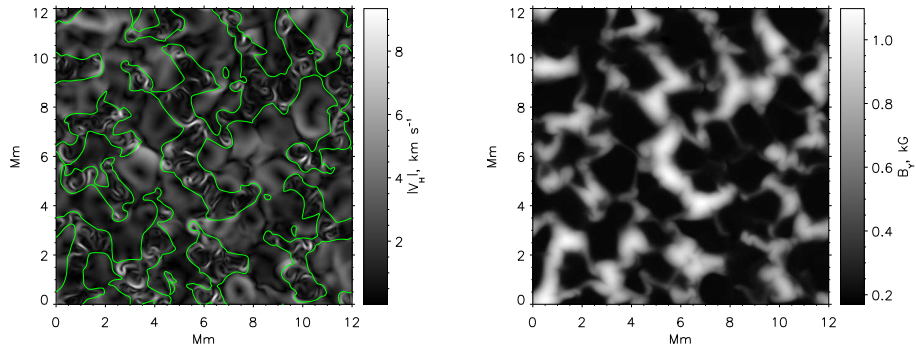}
\caption{Horizontal slices of the modulus of the horizontal components of velocity (left panel) and vertical component of magnetic
field (right panel) measured at the height $y=500~\mathrm{km}$ above the average continuum formation level in the simulation 
domain. Regions with the vertical magnetic field strength larger than $400~\mathrm{G}$,  bounded
by contours on the left panel, are covered by small-scale vortex structures, visually different from granular outflows.}
\label{fig1}
\end{figure}

In Figure~\ref{fig1}, the horizontal slices of the modulus of horizontal components of velocity $v_h=\left(v_x^2+v_z^2\right)^{1/2}$ and the 
vertical component of the magnetic field $\mathrm{B}_y$ are shown in the left and right panels, respectively (note 
that in our notation $x$ and $z$ are the horizontal directions, while $y$ is the vertical). The 
contours on the left panel bound the regions with $\mathrm{B}_y > 400\mathrm{G}$, corresponding to the 
magnetised intergranular lanes. It is evident from the figure that the intergranular vortex motions, visible as 
small-scale semi-circular structures in the left panel of the plot, are bounded by the magnetic field contours, and,
thus are clearly co-located with and abundantly cover the strong intergranular magnetic field regions. Also, there is clearly a 
difference in the geometry of the intergranular magnetic vortex flows and granular
outflows. The granular flows have dark granular centres, corresponding to the upflows with small horizontal
velocities, surrounded by stronger horizontal outflows, which do not show any small-scale structure.

We analyse the small-scale vortex features in the magnetised intergranular lanes, using the vorticity equation,
derived from the equations of magneto-hydrodynamics (MHD). The process of vorticity production was demonstrated by 
\citet{shelyag2011a}. Here we provide only a brief description. The MHD vorticity equation can be derived by 
taking the curl of the MHD momentum equation, separating the magnetic and non-magnetic terms and decomposing 
the magnetic term into two:
\begin{equation}
{\rho}\frac{D}{Dt}\frac{\bf{\omega}}{\rho}=\overbrace{\left(\bf{\omega} \cdot \nabla\right) {\bf v}}^{T_1} -
\overbrace{ \nabla \frac{1}{\rho} \times \nabla p_{g}}^{T_2} - \overbrace{\nabla \frac{1}{\rho} \times 
\left[ \nabla p_m - \left({\bf B} \cdot \nabla \right) {\bf B}\right]}^{T_3} + \overbrace{\frac{1}{\rho}\nabla\times\left[\left({\bf B} 
\cdot\nabla\right){\bf B}\right]}^{T_4}.
\label{voreq1b}
\end{equation}

\begin{figure}
\plotone{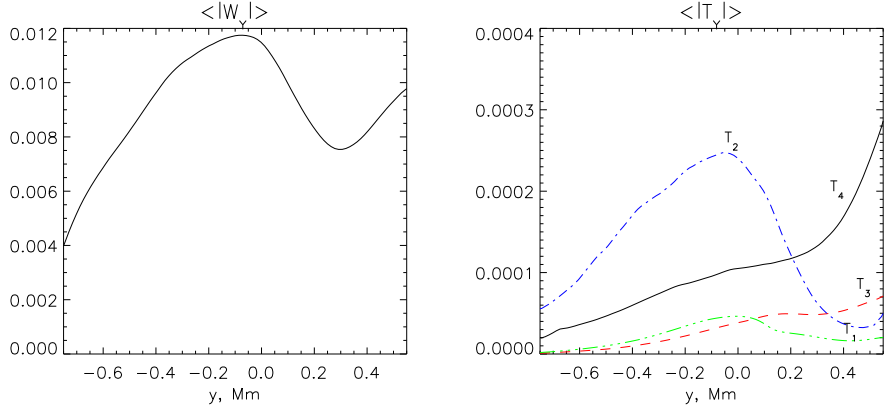}
\caption{Dependence of the horizontally-averaged modulus of the vertical component of vorticity (left panel) and
dependences of the $T_1$ (dash-dot-dotted line), $T_2$ (dash-dotted line), $T_3$ (dashed line) and $T_4$ 
(solid line) terms of vorticity equation (Eq.~\ref{voreq1b}, right panel) on height in the simulation domain. Zero 
height represents the approximate level of the photosphere.}
\label{fig2}
\end{figure}

Here, $T_1$ is the vortex tilting term, $T_2$ the baroclinic vorticity generation term, $T_3$ the magnetic baroclinic term,
and $T_4$ the magnetic field tension term. As shown by \citet{shelyag2011a}, the latter term $T_4$ is mainly
responsible for vorticity production in the upper photosphere, while $T_2$ generates vorticity deeper in the photosphere
and in the convection zone. This is further demonstrated by the right panel of Figure~\ref{fig2}, where the height
dependences of the terms are plotted on the same scale. The figure shows that $T_4$ (solid line) becomes greater
than $T_2$ (dash-dotted line) at heights $0.2~\mathrm{Mm}$ above the approximate continuum formation level, 
while $T_2$ is about twice as large at the level of continuum formation $y=0$. The terms $T_1$ and $T_3$ do 
not play a significant role in the vorticity production. It also should be noted that $T_2$ and $T_4$ act in an opposite way  
and suppress each other, as it is evident from the opposite signs of these terms in Equation~\ref{voreq1b}. In the left
panel of Figure~\ref{fig2}, the dependence of the horizontally averaged modulus of the vertical component of vorticity
is shown. It can be seen from the figure that, in accordance to the physical mechanism outlined above, the vorticity
takes its local minimum at $y=0.2~\mathrm{Mm}$, takes its maximum, caused by the baroclinic term $T_2$ at
$y=0$ and increases towards the upper photosphere in the region $y > 0.2~\mathrm{Mm}$.

\begin{figure}
\plotone{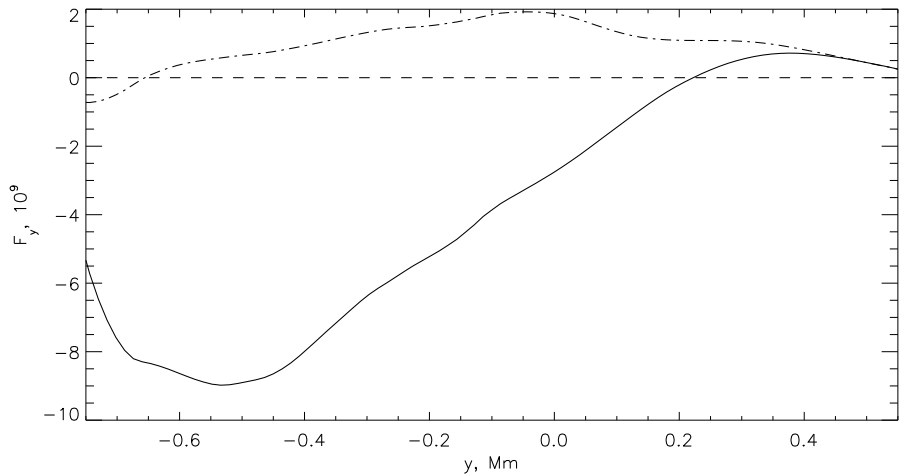}
\caption{Horizontally-averaged dependences of the vertical component of Umov-Poynting flux vector (solid line) and 
the vertical component of the Umov-Poynting flux vector, generated by horizontal plasma motions (dash-dotted line), on 
depth. }
\label{fig3}
\end{figure}

\section{Umov-Poynting flux in the photospheric magnetic vortices}
The plasma motions in magnetic fields can generate Umov-Poynting flux.
In magneto-hydrodynamics, the Umov-Poynting flux vector $\mathbf{F}$ is defined as 
\begin{equation}
\mathbf{F}= \mathbf{B} \times \left(\mathbf{v} \times \mathbf{B}\right),
\label{pflux}
\end{equation}
where $\mathbf{v}$ is the plasma velocity field, and $\mathbf{B}$ is the magnetic field. It is of interest to measure
the vertical, ($y$) component of the Umov-Poynting flux vector ($F_y$), connected to the magnetic vortex motions in the solar 
photosphere, since it can be responsible for the energy transport from the solar interior to the outer solar atmosphere.
Figure~\ref{fig3} shows the dependence of the horizontally-averaged vertical component of the Umov-Poynting flux 
vector on height in the simulated model. As the plot demonstrates, $F_y$ is negative in the convection zone and
in the low photosphere, while it becomes positive (directed outwards) in the upper part of the photosphere above
$y > 0.2~\mathrm{Mm}$, in the same range where the magnetic tension term is greater than the baroclinic hydrodynamic 
term, $T_4 > T_2$ (see Section 2).
%%%
In our simulation we find the behaviour of the vertical component of the Umov-Poynting flux vector to be similar to the result
demonstrated by \citet{steiner1}. However, it should be noted that \citet{steiner1} use a much weaker magnetic field in their simulation, 
and demonstrate a different mechanism of the Umov-Poynting flux generation.

%%%
To confirm the connection between the photospheric magnetic vortices and the Umov-Poynting flux, we analyse the effect
of horizontal motions on the vertical component of the Umov-Poynting flux by setting artificially the vertical component of
velocity $v_y$ to zero in Equation~\ref{pflux}. The depth dependence of $F_{y,v_y=0}$ computed for $v_y=0$ is shown in Figure~\ref{fig3}
as the dash-dotted line. As can be seen, $F_{y,v_y=0}$ is generally positive, except from a small part of the domain
deep in the convection zone, and $F_{y,v_y=0}$ is nearly equal to the full vertical component of the Umov-Poynting flux in the
region $y > 0.4~\mathrm{Mm}$. It is, therefore, clear that the positive vertical component of the Umov-Poynting flux in the upper
photosphere is generated by horizontal vortex motions in the photospheric magnetic field concentrations.

\section{Vortex motions in individual magnetic flux tube embedded into a convectively stable solar atmospheric model}
\begin{figure}
\plottwo{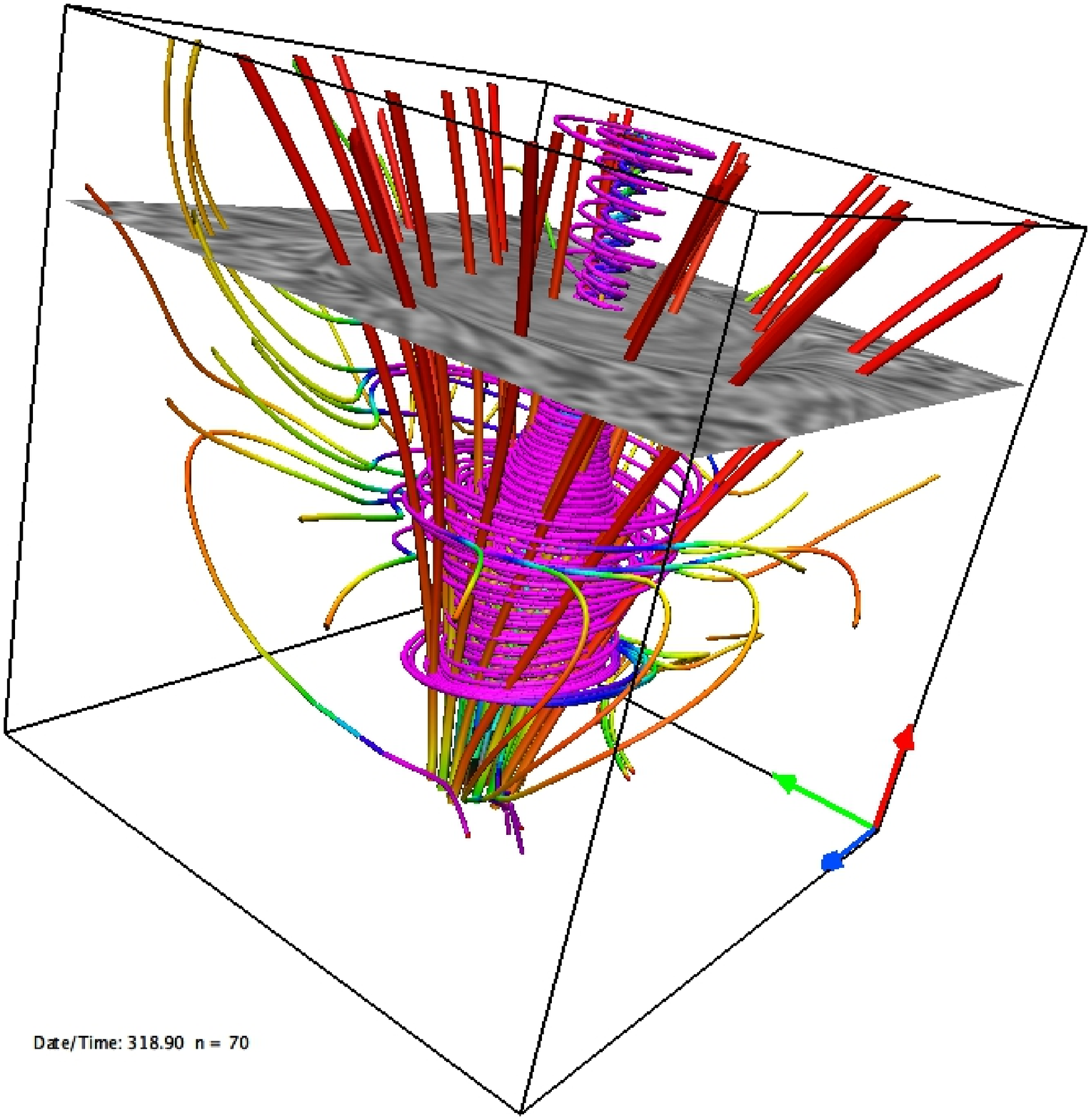}{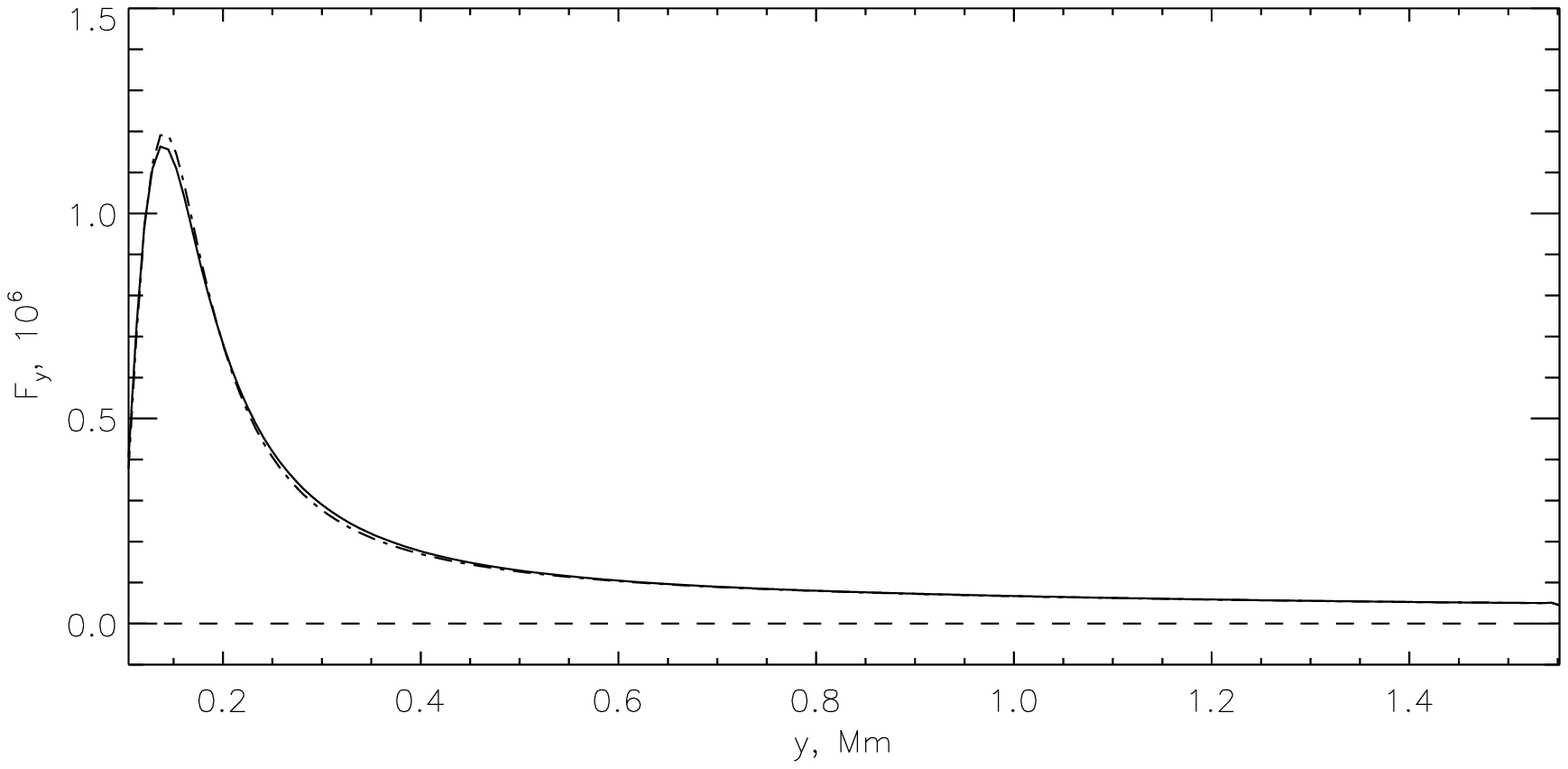}
\caption{Simulation of individual flux tube. Left image - magnetic field lines (red), velocity field lines 
(red-blue-green-violet) and total velocity at $1~\mathrm{Mm}$ height (greyscale) in the domain. Right 
image - vertical component of Umov-Poynting flux.}
\label{fig3a}
\end{figure}

To further establish a relation between photospheric vortex motions in the intergranular magnetic field concentrations
and Umov-Poynting flux, we performed a test simulation of a cylindrically-symmetric magnetic flux tube embedded into
a convectively stable model of the solar atmosphere, using the Sheffield Advanced Code \citep{shelyagcode}. In this simulation,
the computational domain has a physical size of $1.2~\mathrm{Mm^3}$ resolved by 100x100x196 grid points in the $x,~z$ and $y$ 
directions respectively, with a lower boundary  located at the photospheric level $y=0$, and 
magnetic field configuration as described in detail by \citet{fedunan,fedun2011a}. The expanding flux tube with 
the magnetic field strength of $B=1.2~\mathrm{kG}$ at the footpoint (height $y=0.0~\mathrm{Mm}$) is in magneto-hydrostatic
equilibrium with the non-magnetic ambient plasma outside. We perturb the initially stable configuration with a velocity 
source located at height $y=0.1~\mathrm{Mm}$ in the photosphere and acting for the first 200 seconds of the 
simulation. The interaction of the source with the model creates a vortex with the horizontal velocity amplitude 
$v_h \approx 1~\mathrm{km~s^{-1}}$ at the footpoint of the magnetic flux tube (see Figure~\ref{fig3a}; this image was created
with VAPOR 3D visualisation package \citep{vapor}). This vortex 
also generates the vertical Umov-Poynting flux, directed outwards to the upper solar atmosphere. The flux $F_y$ 
is shown in the right panel of Figure~\ref{fig3a}. Similarly to our previous analysis in Section 3, we compare the total 
vertical component of the Umov-Poynting flux with that generated by horizontal motions in the magnetic vortex. The 
two curves representing these are plotted as solid and dash-dotted lines, respectively. Since these are virtually 
indistinguishable, we conclude that the vertical component of Umov-Poynting flux is generated by horizontal vortex motions. 
It should be noted that only a qualitative comparison of this result with the Umov-Poynting flux amplitudes, provided in Section 3, 
is possible due to the different nature of these simulations.

\begin{figure}
\plotone{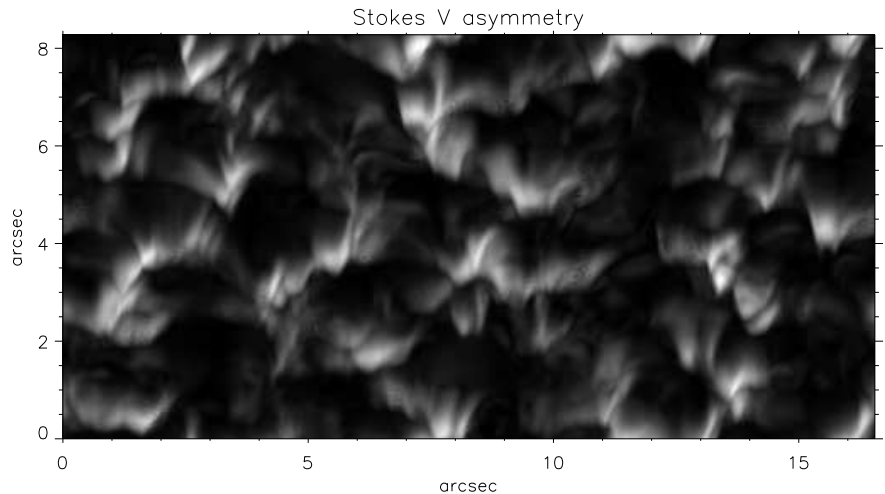}
\caption{Simulated Stokes-$V$ amplitude asymmetry of the FeI $630.2~\mathrm{nm}$ absorption line at $60^{\mathrm{o}}$ 
angular distance to the north from the solar disk centre.}
\label{fig4}
\end{figure}

\section{Radiative diagnostics of the photospheric magnetic vortices}
Intergranular vortices, being horizontal motions in the solar photosphere, are not easy to observe.
The fast changing nature and small size of vortex structures, requires high spatial and temporal resolution.  
Such observations have been used to track spiral motions of photospheric features,  including MBPs  
in the intergranular lanes \citep{bonet1}. However, it is not possible to directly observe the velocity field of 
vortices in the upper photosphere using MBP tracking as the photospheric MBPs are localised intensity 
enhancements which are formed deep in the photosphere and do not fully cover the intergranular lanes.

Therefore, we consider the possibility of directly observing photospheric magnetic vortex 
signatures using spectropolarimetry. Horizontal motions do not reveal themselves in spectropolarimetry at disk centre. 
However, the situation is different near the solar limb as the horizontal
motions show a strong component along the line-of-sight. Also, due to the magnetic nature of the photospheric vorticity,
the asymmetry of Stokes-$V$ profiles \citep[see e.g.][and the references therein]{shelyagasy}, featuring in the regions 
with gradients of line-of-sight magnetic field and velocity
due to layered structure of photospheric vortices \citep{shelyag2011b}, is expected. We analyse this possibility by 
performing radiative transport diagnostics in the photospheric model, inclined by $60~^\mathrm{o}$ to the north
from the solar disk centre, with the FeI $630.2~\mathrm{nm}$ photospheric absorption line. The synthetic Stokes-$V$ 
amplitude asymmetry image, shown in Figure~\ref{fig4}, demonstrates strong signal in the intergranular lanes, as expected. Also, the 
Stokes-$V$ asymmetry features in the image have shapes closely resembling the cylinder-like, layered vortex structures, 
demonstrated by \citet{shelyag2011b}. One of the clearest features of this type is located at $(3,~4)$ arcseconds in 
the image. The spatial resolution required to resolve this feature is estimated to be about $0.1~\mathrm{arcsec}$ in low light 
conditions due to limb darkening. Thus, future instruments for high-resolution spectropolarimetric solar observations
with large apertures, such as the planned ATST, may be able to resolve these features in detail.

\section{Conclusions}
Using direct numerical modelling of solar photospheric plasma, we simulate the vortex motions in the 
magnetised intergranular lanes. Through two different approaches of simulations of magneto-convection, and 
direct modelling of the magnetic flux tube embedded into the solar atmosphere, we demonstrate that magnetised intergranular 
lanes generate significant amount of Umov-Poynting flux, directed outwards to the higher layers of the solar 
atmosphere, and show that the main part of this flux is produced by horizontal vortex motions. Finally, we 
demonstrate the possibility to directly observe photospheric vortices using high-resolution spectropolarimetric 
observations close to the solar limb.

\acknowledgements This work has been supported by the UK Science and Technology Facilities Council (STFC). 
R.~Erd{\'e}lyi acknowledges M. K{\'e}ray for patient encouragement and is grateful to NSF, Hungary (OTKA, Ref. 
No. K83133). S.~Shelyag is grateful to ATST-EAST workshop organisers for providing help and financial support 
to attend the meeting. S.~Shelyag is also grateful to R. Stein and M. Carlsson for helpful discussions during the meeting.

\thebibliography

\bibitem[Bonet et al.(2008)]{bonet1} Bonet, J.~A., M{\'a}rquez, I., S{\'a}nchez Almeida, J., Cabello, I., \& Domingo, V.\ 2008, \apjl, 687, L131 

\bibitem[Clyne et al.(2007)]{vapor} Clyne, J., Mininni, P., Norton, A., \& Rast, M.\ 2007, New Journal of Physics, 9, 301 

\bibitem[Erd{\'e}lyi \& Fedun(2007)]{erdelyi1} Erd{\'e}lyi, R., \& Fedun, V.\ 2007, Science, 318, 1572 

\bibitem[Fedun et al.(2011a)]{fedunan} Fedun, V., Shelyag, S., Verth, G., Mathioudakis, M., \& Erd{\'e}lyi, R.\ 2011, Annales Geophysicae, 29, 1029 

\bibitem[Fedun et al.(2011b)]{fedun2011} Fedun, V., Shelyag, S., \& Erd{\'e}lyi, R.\ 2011, \apj, 727, 17

\bibitem[Fedun et al.(2011c)]{fedun2011a} Fedun, V., Verth, G., Jess, D.~B., \& Erd{\'e}lyi, R.\ 2011, \apjl, 740, L46

\bibitem[Jess et al.(2009)]{jess1} Jess, D.~B., Mathioudakis, M., Erd{\'e}lyi, R., et al.\ 2009, Science, 323, 1582 

\bibitem[Moll et al.(2011)]{moll1} Moll, R., Cameron, R.~H., \& Sch{\"u}ssler, M.\ 2011, \aap, 533, A126

\bibitem[Rempel et al.(2009)]{rempel1} Rempel, M., Sch{\"u}ssler, M., Cameron, R.~H., \& Kn{\"o}lker, M.\ 2009, Science, 325, 171 

\bibitem[Shelyag et al.(2004)]{shelyagbp2} Shelyag, S., Sch{\"u}ssler, M., Solanki, S.~K., Berdyugina, S.~V.,  V{\"o}gler, A.\ 2004, \aap, 427, 335 

\bibitem[Shelyag et al.(2007)]{shelyagasy} Shelyag, S., Sch{\"u}ssler, M., Solanki, S.~K., V{\"o}gler, A.\ 2007, \aap, 469, 731 

\bibitem[Shelyag et al.(2008)]{shelyagcode} Shelyag, S., Fedun, V., \& Erd{\'e}lyi, R.\ 2008, \aap, 486, 655 

\bibitem[Shelyag et al.(2011a)]{shelyag2011a} Shelyag, S., Keys, P., Mathioudakis, M., \& Keenan, F.~P.\ 2011, \aap, 526, A5 

\bibitem[Shelyag et al.(2011b)]{shelyag2011b} Shelyag, S., Fedun, V., Keenan, F.~P., Erd{\'e}lyi, R., \& Mathioudakis, M.\ 2011, Annales Geophysicae, 29, 883 

\bibitem[Steiner et al.(2008)]{steiner1} Steiner, O., Rezaei, R., Schaffenberger, W., \& Wedemeyer-B{\"o}hm, S.\ 2008, \apjl, 680, L85

\bibitem[Vargas Dom{\'{\i}}nguez et al.(2011)]{vargas1} Vargas Dom{\'{\i}}nguez, S., Palacios, J., Balmaceda, L., Cabello, I., 
\& Domingo, V.\ 2011, \mnras, 416, 148 

\bibitem[V{\"o}gler et al.(2005)]{voegler1} V{\"o}gler, A., Shelyag, S., Sch{\"u}ssler, M., Cattaneo, F., Emonet, T., \& Linde, T.\ 2005, \aap, 429, 335 

\bibitem[Wedemeyer-B{\"o}hm \& Rouppe van der Voort(2009)]{wedemeyer1} Wedemeyer-B{\"o}hm, S., \& Rouppe van der Voort, L.\ 2009, \aap, 507, L9 

\bibitem[Zaqarashvili et al.(2011)]{zaqarashvili1} Zaqarashvili, T.~V., Khodachenko, M.~L., \& Rucker, H.~O.\ 2011, \aap, 534, A93 

\end{document}